\documentclass[useAMS,usenatbib]{mn2e}
\usepackage{graphicx}
\usepackage{txfonts}
\title[IGR J14488-4008: an X-ray peculiar giant radio galaxy discovered by INTEGRAL]
  {IGR J14488-4008: an X-ray peculiar giant radio galaxy discovered by INTEGRAL}
\author[M. Molina et al.]
{M. Molina$^1$, T. Venturi$^2$, A. Malizia$^1$, L. Bassani$^1$, D. Dallacasa$^{2,3}$, D. Vir Lal$^4$, A.J. Bird$^5$,
\newauthor
P. Ubertini$^6$ \\
$^1$IASF/INAF, via Gobetti 101, I-40129 Bologna, Italy\\
$^2$IRA/INAF, via Gobetti 101, I-40129 Bologna, Italy\\
$^3$Dipartimento di Fisica e Astronomia, Universit$\grave{a}$ di Bologna, Via Ranzani 1, 40127, Bologna, Italy\\
$^4$National Centre for Radio Astrophysics, Pune Univ. Campus, Ganeshkhind, Pune 411 007, India\\
$^5$School of Physics and Astronomy, University of Southampton, SO17 1BJ, Southampton, UK\\
$^6$IAPS/INAF, Via Fosso del Cavaliere 100, I-00133 Roma, Italy}
\begin{document}

\date{}

\pagerange{\pageref{firstpage}--\pageref{lastpage}} \pubyear{2014}

\maketitle

\label{firstpage}
        
\begin{abstract}
In this paper we report the discovery and detailed radio/X-ray analysis of a peculiar giant radio
galaxy (GRG) detected by INTEGRAL, IGR J14488--4008. The source has been recently classified
as a Seyfert 1.2 galaxy at redshift 0.123; the radio data denote the source to be a type II
Fanaroff-Riley radio galaxy, with a linear projected size exceeding 1.5\,Mpc, clearly assigning IGR J14488--4008
to the class of GRG. In the X-rays, the source shows a remarkable spectrum, characterised by 
absorption by ionised elements, a characteristic so far found in only other four broad line radio galaxies.
\end{abstract}


\section{Introduction}

Giant radio galaxies (GRGs) are the largest and most energetic single objects in the Universe 
and are of particular interest as extreme examples of radio source development and evolution. 
With linear sizes exceeding 0.7\,Mpc\footnote{Scaled
for a cosmology with H$_0$=71\,km\,s$^{-1}$\,Mpc$^{-1}$, $\Omega_{\rm m}$=0.27,
$\Omega_{\Lambda}$=0.73; this value is scaled from the original 1\,Mpc to compensate for recent  
changes in the cosmological constants.} (e.g. \citealt{lara:2001};
\citealt{Ishwara-Chandra:1999}), they represent the oldest and long-lasting part of the radio galaxies population: 
assuming that the spectral ages of radio galaxies are representative of their dynamical ages 
\citep{parma:1999}, GRGs result to have on average radiative ages in excess of 10$^8$ years. 
Despite this, GRGs display bright radio features which are indicative of current or recent activity 
of the central engine. Understanding how they can manage to maintain an AGN active for so long 
and capable of producing radio emission for a long time (while the nucleus in most galaxies 
becomes quiescent in a much shorter period, see e.g. \citealt{schoenmakers:2000} and \citealt{parma:1999}) 
requires as much data as possible on a fairly 
extended sample of giant radio galaxies. 

An interesting new development in the study of GRGs 
has recently come from the detection of a substantial fraction of these objects among soft gamma-ray 
selected radio galaxies. Since 2002, the soft gamma-ray sky has been surveyed by 
INTEGRAL/ISGRI and Swift/BAT at energies higher than 20\,keV; up to now, various all sky 
catalogues have been released (see for example, \citealt{Bird:2010}; \citealt{Baumgartner:2013}) 
revealing a large population of active galaxies. Around 7\% of these soft gamma-ray selected AGN 
display a double-lobed morphology typical of a radio galaxy. What is intriguing and 
particularly interesting is that a large fraction ($\sim$30\%) of these radio galaxies have 
giant radio structures (Bassani et al. in preparation). 

The fraction of GRGs 
in radio source samples is generally small ($\sim$6\% in the 3CR catalogue, \citealt{Ishwara-Chandra:1999}), 
and it becomes relevant among radio loud soft gamma-ray selected objects. Therefore it suggests a  
link between the central AGN properties and the extended radio structures. To date, 
about 20 objects within the INTEGRAL and Swift surveys have been identified as GRGs; 
a few of these are new discoveries like IGR J17488--2338 \citep{molina:2014} and IGR J14488-4008,  
the source discussed in the present paper. These 
findings confirm that observations at soft gamma-rays are useful tools to 
find radio galaxies with giant structures but also suggests that combining radio observations with 
measurements at high energies is a new and useful approach to study GRG.

Here we report on the discovery and consequent radio and X-ray
analysis of the INTEGRAL source IGR J14488--4008,
recently classified as a Seyfert 1.2 galaxy at redshift z=0.123.
The radio data indicate that the source has
a Fanaroff-Riley FR\,II  morphology \citep{fanaroff:1974},
and linear size exceeding 1.5\,Mpc, making it a newly discovered giant radio galaxy. 
In the X-ray, the source shows a peculiar spectrum, characterised by ionised absorption,
a characteristic which has so far been found only in four other broad line radio galaxies.

\section{High Energy Discovery and Identification}
\begin{figure}
\centering
\includegraphics[width=0.85\linewidth]{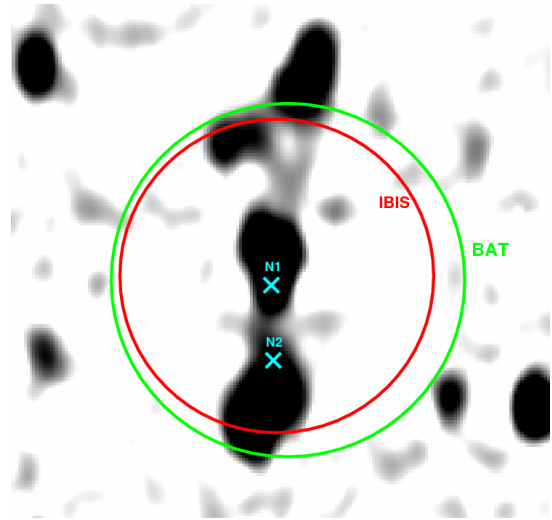}
\caption{843\,MHz SUMSS radio image of the region containing IGR J14488--4008. The larger, green
circle marks the BAT position and relative uncertainty (5.07 arcmin); the red circle marks instead
the ISGRI position and its uncertainty (4.5 arcmin). The two objects marked as N1 and N2 are the
two X-ray sources detected by XRT. N1 is the true counterpart to the high energy source. }
\label{sumss}
\end{figure}

IGR J14488--4008 was first reported as a hard X-ray emitter 
in the 4$^{\rm th}$ INTEGRAL IBIS catalogue \citep{Bird:2010} and located, 
with a positional uncertainty of 4.5 arcmin,
at RA(J2000) = 14h 18m 50.2s and Dec(J2000) = -40d 08m 31s. 
The source is reported in the \emph{Swift}-BAT 70-month catalogue
and named SWIFT J1448.7--4009.
In order to reduce the positional uncertainty associated with the
INTEGRAL detection and therefore to favour identification, 
in 2011 the source was the subject of a follow-up campaign carried on 
with \emph{Swift}-XRT \citep{malizia11}. Within the ISGRI/BAT error circles, XRT detected
two X-ray sources (see Figure \ref{sumss}): 1) a bright source (N1),
located at RA(J2000) = 14h 48m 50.98s and
Dec.(J2000) = -40d 08m 47.12s (with a positional uncertainty of 3.83 arcsec)
and with an X-ray flux of 2.4$\times$10$^{-12}$ erg\,s$^{-1}$\,cm$^{-2}$
(see \citealt{malizia11} and \citealt{molina:2012}); 2) a dimmer
object (N2) located at RA(J2000) = 14h 48m 50.82s and Dec(J2000) = -40d 10m 56.70s, 
with a very soft spectrum (undetected
above 3\,keV). 
As evident in Figure \ref{sumss}, which is a SUMSS (Sydney University Molonglo Sky 
Survey; \citealt{bock99}) image of the sky region around IGR J14488--4008,
both objects appear overlaid to
a bright and complex radio source, showing a nuclear region plus a two-sided lobe structure.
As reported by \citet{molina:2012}, this radio source was observed at several radio frequencies: 
at 1.4\,GHz with a flux of 48\,mJy (NVSS; \citealt{condon98}); at 4.85\,GHz with a flux of 
48\,mJy (Combined Radio All-sky Targeted Eight GHz Survey-CRATES; \citealt{healey07}); 
at 8.4\,GHz with a flux of 45.9\,GHz  and at 8.6\,GHz with a flux of 46\,mJy 
(ATPM, \citealt{mcconnell12}). 
IGR J14488--4008 was also observed during the course of an optical follow-up programme by
\citet{masetti:2013}; these authors found source N1 to be a Seyfert 1.2 galaxy at z = 0.123,
with a black hole mass of (3.8$\pm$1.9)$\times$10$^8$M$_{\odot}$ and
source N2 to be a G-type star with no peculiarities and unlikely  
to produce high energy emission.

As can be seen from Figure \ref{sumss}, the resolution of the 843\,MHz SUMSS 
image is not adequate to clearly identify the various components of the radio structures
that characterise IGR J14488--4008 in this band and also is unable to associate source
N1 to the core of the radio object.
IGR J14488--4008 belongs to a sample of giant radio galaxies detected by INTEGRAL
which is currently under study with the Giant Metrewave Radio
Telescope (GMRT) (see \ref{radio}). To further study the source in the X-ray band, we also
carried out an XMM-\emph{Newton} observation in AO12.

\section{Radio observations and imaging}\label{radio}
 The GMRT observations were performed at 610\,MHz (on 9 June 2014) and at 
325\,MHz (on 14 May 2014), for a total of 5 hours at each frequency. The 
observing bandwidth at each frequency is 64\,MHz, split into 256 channels
with $\Delta\nu$=250\,kHz.
The starting steps of the data reduction, i.e. RFI removal and bandpass 
calibration, were done using {\it flagcal} \citep{prasad:2012}, while self-calibration and
imaging were carried out using the NRAO Astronomical Image Processing System 
(AIPS). We performed self-calibration using all the sources in the GMRT
field of view at each frequencies.
At both frequencies we produced final images at full and low resolution, 
to properly image both compact and extended features of the radio galaxy.
The noise in the final images is in the range 
$1\sigma\sim$0.12--0.20\,mJy b$^{-1}$ at\,325 MHz and 
$1\sigma\sim 40-80 \mu$Jy b$^{-1}$ at 610\,MHz, depending on the resolution
and on the position in field. 

\begin{figure}
%\centering
\includegraphics{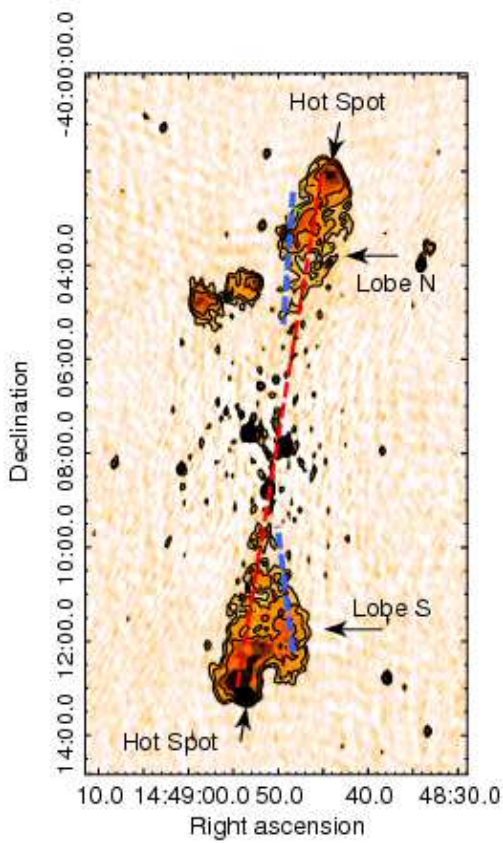}
\caption{Radio 610\,MHz GMRT full resolution
image of IGR J14488--4008 with radio contours (at the same frequency) overlaid; the restoring beam
is 8.77$^{\prime}$$^{\prime}$$\times$4.92$^{\prime}$$^{\prime}$ with a rms of 30\,mJy/beam around the
source edges and of 50\,mJy/beam close to the source. }
%\emph{Right panel}: 2--10\,keV XMM EPIC-pn image of the source with the 610\,MHz radio contours
%overlaid. }
\label{fig:overlay}
\end{figure}

\begin{figure}
\centering
\includegraphics[width=0.85\linewidth]{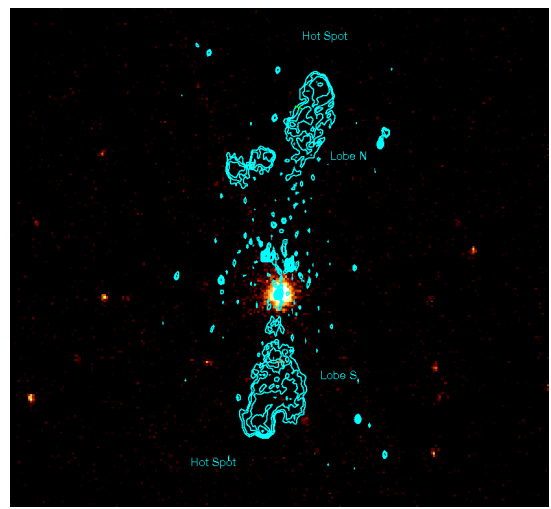}
\caption{0.3-10\, keV XMM-Newton image of IGR J14488--4008, with the 610\,MHz radio contours
overlaid as in figure \ref{fig:overlay}.}
\label{xmm_radio}
\end{figure}

 Figure \ref{fig:overlay} shows the 610\,MHz GMRT contours overlaid on the
same frequency GMRT image. The high resolution and sensitivity of the GMRT images 
reveal important details on the radio emission of IGR J14488--4008, compared
to the image provided by SUMSS.
The central component in SUMSS is actually a blend of three compact components.
The overlay of the radio image with the XMM-Newton image (see Figure \ref{xmm_radio})
clearly shows that the core of the radio
galaxy is the southernmost one, and it is coincident with the
X-ray source labelled N1 in Figure \ref{sumss}. None of the other two compact
components is optically identified on the red plate of the red
Digitized Sky Survey DSS2, and the available data (in the optical
and in the radio) do not suggest that these two components are
the lobes or hot spots of a FR\,II background radio source. Based on
our GMRT images their spectral index
$\alpha_{\rm 322~MHz}^{\rm 607~MHz}$ is 0.5, typical of compact
radio sources.
Moreover, the eastern extension of the northern lobe, as visible on
SUMSS, is actually a background radio galaxy with FR\,II morphology (no
redshift is available in the literature for the optical counterpart).

A noticeable feature of the large scale structure of IGR J14488--4008
is the misalignment in the morphology of the lobes. In particular,
the high surface brightness regions of the lobes (including the hot spots)
are symmetric with respect to the core, and are aligned in position
angle of $\sim$-12$^{\circ}$, while the low surface brightness parts are
symmetric with respect from the core but their alignment is
$\sim$-5$^{\circ}$. This is suggestive of a X--shaped radio galaxy.
If confirmed, IGR J14488--4008 would be the  second giant X--shaped
radio galaxy \citep{saripalli:08}.

Table \ref{tab:radio} reports the flux density of each component at the 
325\,MHz and 610\,MHz, and the spectral index between these two frequencies.
The total radio power is typical for FR\,II radio galaxies, 
LogP$_{\rm 325~MHz}$=25.36 W Hz$^{-1}$, 
and the largest linear size of IGR J14488--4008, estimated as the projected 
distance between the peaks of the hot spots is 1.54\,Mpc. This size fully
qualifies IGR J14488--4008 as a giant radio galaxy.
The core is inverted in this frequency range, and the integrated 
spectrum of the source is flatter than the canonical values 0.7-0.8 
(assuming S$_{\nu}$$\propto$$\nu^{-\alpha}$)
usually reported for FR\,II radio galaxies. This is most likely due to the fact that the
flux density of the radio galaxy is dominated by the hot spots and the
beginning of the backflow of the lobes. We point out that the frequency
range of the observations presented here is below 1 GHz, where the
spectrum of the radio galaxy may still be dominated by the initial
population of relativisti electrons, prior to ageing.
We estimated the intrinsic physical parameters of the radio lobes under
equipartition assumptions (filling factor $\Phi$=1 and energy ratio 
between protons and electrons $k=1$), and the values are reported in Table \ref{tab:physical}.
The two lobes are very similar, and the equipartition parameters are
considerably lower (at least one order of magnitude) compared to radio
galaxies of smaller size\footnote{Such values have been derived under the usual assumption of 
identical energy in electrons and protons, and a volume homogeneously filled with relativistic plasma.}, 
and are lower than found in the two other giant
radio galaxies 3C\,35 and 3C\,223 under similar assumptions
\citep{orru:2010}. 

The lack of information on the spectral index shape
and high-frequency cutoff of the two lobes in IGR J1448--4008 does not
allow an estimate of the radiative age of the source.

\begin{table}
\caption[]{The radio images}
\begin{center}
\begin{tabular}{lccc}
\hline
\hline
%1                   &    2  & 3 &      4  \\
Component & S$_{\rm 325~MHz}$ & S$_{\rm 610~MHz}$ & $\alpha_{\rm 325~MHz}^{\rm 610~MHz}$ (a)\\ 
          &  mJy              &     mJy           &  \\
\hline
Core    &   20.7$\pm$ 0.8 &   32.3$\pm$ 1.3 & -0.71$\pm$0.14 \\
Lobe N  &  201.9$\pm$10.1 &  148.1$\pm$ 7.7 &  0.49$\pm$0.16 \\
Lobe S  &  354.4$\pm$17.7 &  263.0$\pm$13.2 &  0.47$\pm$0.16 \\
Total   &  576.9$\pm$28.8 &  443.4$\pm$22.2 &  0.42$\pm$0.15 \\

\hline
\end{tabular}
\end{center}
\label{tab:radio}
Notes: (a) The sign convention is S$\propto\nu^{-\alpha}$.
\end{table}

\begin{table}
\caption[]{Physical parameters}
\begin{center}
\begin{tabular}{lccc}
\hline
\hline
Component & u$_{\rm min}$ & P$_{\rm min}$ & B$_{\rm eq}$\\ 
%          &  erg cm$^{-3}$ &     dyn cm$^{-2}$& $\mu$G \\
\hline
Lobe N  &  4.5$\times$10$^{-14}$erg cm$^{-3}$ &  2.4$\times$10$^{-14}$dyn cm$^{-2}$ &  0.7$\mu$G  \\
Lobe S  &  5.5$\times$10$^{-14}$erg cm$^{-3}$ &  2.9$\times$10$^{-14}$dyn cm$^{-2}$&  0.8$\mu$G \\
\hline
\end{tabular}
\end{center}
\label{tab:physical}
\end{table}

\section{X-ray Timing and Spectral Analysis}

%\subsection{XMM-\emph{Newton} observation}\label{xobs}
IGR J14488--4008 was also recently observed by XMM-\emph{Newton} on Feb. 16, 2014, 
during revolution 2599.
The EPIC-pn \citep{Turner:2001} data were reprocessed using the \emph{XMM-Newton} Standard
Analysis Software (SAS) version 12.0.1 and employing the latest available calibration files. 
Only patterns corresponding to single and double events (PATTERN$\leq$4) were taken
into account; the standard selection filter FLAG=0 was applied.  
The EPIC-pn nominal exposure of 28\,ks was filtered for periods of high background, resulting
in a cleaned exposure of $\sim$17.6\,ks. Source counts were extracted from 
a circular region of 29.5$^{\prime\prime}$ radius centred on the source, 
while background spectra were extracted from two circular regions of 20$^{\prime\prime}$ of 
radius close to the source. The ancillary response matrix (ARF) and the detector 
response matrix (RMF) were generated using the \emph{XMM}-SAS tasks \emph{arfgen} and 
\emph{rmfgen} and spectral channels were rebinned in order to achieve a 
minimum of 20 counts per bin. No pile-up was detected for this source. 
The acquisition of the XMM data allowed us to put firm constraints on the 
source identification with the bight radio galaxy; it also allowed us to exclude
X-ray emission from either the lobes or hot spots of the GRG.

As a preliminary step in the analysis of IGR J14488-4008, we focused on the issue
of variability, both on short time scales, through the analysis of the XMM light curves, 
and on longer time scales, employing light curves from ISGRI/ISGRI and Swift/BAT.

The XMM light curves were produced in three different energy bands 
(soft, 0.3--2\,keV; hard, 2--10\,keV; total, 0.3--10\,keV) and then analysed with
the \texttt{XRONOS} package. The statistical analysis of the light curves in the
chosen energy bands, does not point to any significant evidence of short term variability, as
shown by the relative $\chi^2$. In fact, for a 100s binned light curve, we find
for the soft band (0.3--2\,keV) light curve a $\chi^2$/d.o.f of 112.7/165, 
for the hard band light curve (2--10\,keV), we find a
$\chi^2$ of 121.9/163 and finally for the 0.3--10\,keV
light curve we find $\chi^2$=120.9/164. 
For the three light curves we find a probability of constancy of
more than 99\%.
Also the analysis of the soft vs. hard light curve hardness ratio excludes any kind
of spectral variability. 

We checked as well the hard X-ray light curves
obtained by IBIS/ISGRI and Swift/BAT to examine if long-term variability is instead present;
also in this case, we found that variability on long time-scales (months) is not significant 
and could be accounted for by introducing a cross-calibration constant between instruments when 
fitting the broad-band spectrum, as discussed in section \ref{broad_band}.

Spectral analysis was performed using \texttt{XSPEC} v.12.8.0 \citep{Arnaud:1996}; errors are 
quoted at 90\% confidence level for one parameter of interest ($\Delta\chi^2$=2.71). 
In the following analysis, together with the proprietary XMM observation, 
we use data collected in the fourth IBIS survey \citep{Bird:2010}, 
which consists of all exposures from the beginning of the mission 
(November 2002) up to April 2008 (revolutions 12 through 530). The IBIS/ISGRI 
data analysis and the average source spectra extraction has been obtained following the 
procedure described by \citet{Molina:2013}. To enhance the available statistics,
above all at high energies, we also make use of the 
accessible Swift/BAT spectrum \citep{Baumgartner:2013} of IGR J14488--4008,
obtained from the latest 70-month 
catalogue\footnote{http://swift.gsfc.nasa.gov/results/bs70mon/}. 
XMM data have been fitted in the 0.3--10\,keV energy range, while the ISGRI and BAT 
spectra have been fitted in the 20--110\,keV and 14--110\,keV bands
respectively. Soft and hard X-ray data have then been combined together in order to
obtain a broad-band coverage in the 0.3--110\,keV energy range.

\subsection{The soft 0.3-10 keV XMM-Newton spectrum}\label{xmm_analysis}

\begin{figure}
\centering
\includegraphics[scale=0.3, angle=-90]{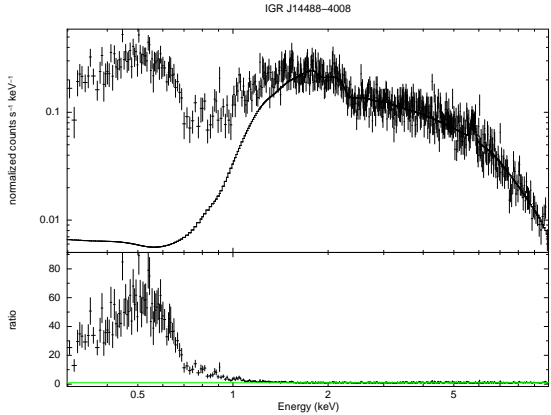}
\caption{0.3--10\,keV EPIC-pn spectrum and residuals,; the model employed is
\texttt{wa$_{\rm g}$*zwa*(zpo+zga)}. As it is evident, a prominent
soft excess component is present below 1\,keV.}
\label{xmm_simple_pl}
\end{figure}

The EPIC-pn data were initially fitted in the restricted 2--10\,keV energy range,
in order to first characterise the primary continuum. 
As a first step, we fitted the data with a simple power law absorbed only by the Galactic column
column density (\texttt{wa$_{\rm g}$*zpo} in \texttt{XSPEC}, with
N$_{\rm H}^{\rm Gal}$=0.07$\times$10$^{22}$cm$^{-2}$); as expected,
the fit is very poor ($\chi^2$=457.90 for 379 d.o.f.) as some additional features (such
as a cold absorption and an iron line component) could be present.
We therefore added first a cold absorption component to our model 
(\texttt{wa$_{\rm g}$*zwa*zpo} in \texttt{XSPEC});
the fit improves ($\chi^2$=403.92 for 378 d.o.f.) and the component is highly required
by the data (at more than 99\% confidence level). We then included a Gaussian component to the fit
(\texttt{wa$_{\rm g}$*zwa*(zpo+zga)}): the fit further improves ($\chi^2$=380.20, 375 d.o.f.)
and the new component is required at more than 99\% coincidence level. 
The iron line is indeed well detected at 6.45\,keV and likely to be broad 
(we have an upper limit on the line width at 139\,eV and an equivalent width of 102\,eV).
After characterising the spectrum in the 2--10\,keV energy range, we extended our analysis
to the softer energies. As shown in Figure \ref{xmm_simple_pl},
evidence of a strong soft excess is visible below 2\,keV.
 As a first step towards the characterisation of this soft excess, we analysed
the data using standard thermal models. When a blackbody component 
(\texttt{wa$_{\rm g}$*zwa*(zbb+zpo+zga)}) is employed to fit our data,
the resulting $\chi^2$ is quite poor (1101.96 for 588 d.o.f.) and the parameter values are
inconsistent with those expected for AGN. We subsequently substituted the blackbody component
with the \texttt{XSPEC} model \texttt{mekal}, but, although the fit is better ($\chi^2$= 699.27, 588 d.o.f.),
the spectrum still shows residuals at soft energies, where absorption edges might be present.
We also tested if cool Comptonisation or reflection models could properly fit the data, but in both cases
we did not obtain acceptable fits. In the first case, we tried to fit the soft excess by employing the
\texttt{compTT} model, while the continuum was modelled with a power law plus a reflection
component due to the accretion disc (local model \texttt{reflionx}), intrinsic cold absorption and 
a Gaussian component to fit the Fe K$\alpha$ line. The fit
was however poor, yielding a $\chi^2$ of 2468.62 (for 586 d.o.f.). We then substituted the cool
Comptonisation model with a reflection model, consisting in a relativistically smeared
reflector (the final model was \texttt{wa$_{\rm g}$*(kdblur(zpo+reflionx)+zga)} in \texttt{XSPEC}
terminology), but again the fit was poor, having a $\chi^2$ of 2037.42 (588 d.o.f.).
Therefore, we investigated on the possibility 
that the observed soft excess could be an artefact of smeared absorption from partially ionised 
material as described by \citet{gierlinsky:2004}. To test this scenario, we have fitted the 
EPIC-pn data with the smeared absorption from a partially ionised wind from an accretion disk 
(\texttt{swind1} in \texttt{XSPEC}) in addition to the absorbed power-law model for the continuum 
and the Gaussian line for the Fe k$\alpha$ line. The parameters of the smeared wind model are 
the absorbing column density, the ionisation parameter
($\xi$ = L/nr$^2$)\footnote{In the formula L is the source luminosity, n is the density and r the 
distance between the ionising source and the absorbing gas} 
and the Gaussian sigma for velocity smearing in units of v/c. Although it has been shown 
that this model requires non physical, unusually 
high terminal velocities of the outflow  \citep{schurch:2008}, 
it can still be used as a proxy for multiple-absorber, 
partially-covering, ionised absorber models that can account for the soft excess.
The smeared absorption model provided a good fit to the spectrum, yielding a
$\chi^2$ of 759.48 for 587 d.o.f. and the new component is highly required by the data
(at more than 99\% confidence level). In the smeared wind model framework, lines from atomic recombination are expected
to be found in the source spectrum. Indeed, although the fit is acceptable, inspection of 
the residuals show the presence of absorbing features at around 0.7 and 0.9\,keV.
Hence, we tested our data for the presence of the O$_{\rm VII}$ and
Fe$_{\rm XX}$ absorption edges, located at around 0.73\,keV and 0.96\,keV respectively. 
The O$_{\rm VII}$ absorption edge at 0.73\,keV  has 
$\tau_{\rm max}$ of 2.58$^{+0.55}_{-0.40}$, while the Fe$_{\rm XX}$ absorption edge 
at 0.96\,keV has $\tau_{\rm max}$=0.71$^{+0.36}_{-0.31}$; both components
are required by the data at more than 99\% confidence level.
The 2--10\, keV flux and luminosity are 5.26$\times$10$^{-12}$erg\,cm$^{-2}$\,s$^{-1}$
and 1.93$\times$10$^{44}$erg\,s$^{-1}$ respectively.
To conclude, we find that the 0.3Ð-10\,keV best-fit model to the XMM data is
\texttt{wa$_{\rm g}$*zwa*zedge*zedge*swind1*(zpo+zga)}.

\subsection{Broad-Band Spectral Analysis}\label{broad_band}

Extrapolation of the XMM best fit model to the 20--110\,keV band provides a slightly higher 
flux than those obtained by fitting the ISGRI and BAT spectra individually.
This suggests that long term flux variability could be present, although one must consider 
the possibility of a more complex spectrum when high energy data are considered. The long term 
variability can be accounted for using a cross-calibration constant between instruments
when fitting the broad-band (0.3--110\,keV) data. When this is done using the XMM best-fit model, 
the fit is acceptable yielding a $\chi^{2}$ of 622.68 
(604 d.o.f.) and the cross-calibration constants between XMM/BAT and XMM/ISGRI are 0.75 $\pm$0.21 
and 0.55$\pm$0.24 respectively.
However, inspection of the residuals indicates a curvature in the spectrum above 20\,keV,
suggesting that some additional components might be required at
higher energies to better represent the source spectrum. The presence of a high energy cut-off 
was consequently tested by substituting the simple power law in the baseline model with
an exponentially cut-off power law reflected from neutral material 
(\texttt{wa$_{\rm g}$*zwa*zedge*zedge*(pexrav+zga)} in \texttt{XSPEC} terminology).
The reflection component was added to account for the presence of the observed characteristics of
the iron line; due to the statistical quality of the high energy data,
we fixed the reflection fraction to 1 given the value of the Fe line 
EW (since EW/R$\sim$100-130\,eV; \citealt{perola:2002}.) 
The fit is acceptable, yielding a $\chi^2$ of 772.76 for 605 d.o.f. 
the power law slope is slightly steeper than the one found for the soft X-ray fit alone, 
being $\Gamma$=1.76$^{+0.09}_{-0.10}$, and the high energy 
cut-off is unfortunately not very well constrained (E$_{\rm cut}$=75$^{+398}_{-45}$\,keV).
The O$_{\rm VII}$ and Fe$_{\rm XX}$ absorption edges energies have been fixed at the values found
for the soft X-ray fit, at 0.73\,keV and 0.96\,keV respectively, while their depth has been left 
as a free parameter. The O$_{\rm VII}$ absorption edge at 0.73\,keV  has 
$\tau_{\rm max}$ of 2.58$^{+0.55}_{-0.40}$, while the Fe$_{\rm XX}$ absorption edge 
at 0.96\,keV has $\tau_{\rm max}$=0.71$^{+0.36}_{-0.31}$.% both components
%are required by the data at more than 99\% confidence level. 

\begin{figure}
\centering
\includegraphics[scale=0.3, angle=-90]{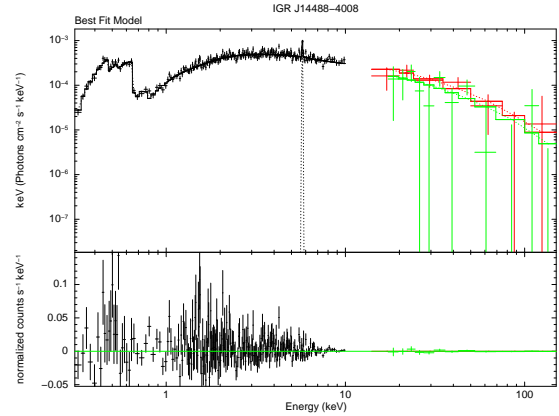}
\caption{0.3--110\,keV XMM/BAT/ISGRI best-fit plot. Black data points are the XMM 
measurements, green data points are the BAT spectrum and red data points are relative 
to the ISGRI one.}
\label{bb_eufspec}
\end{figure}

The fit we obtain is our best fit, with a $\chi^2$ of 614.99 for 603 d.o.f ($\Delta\chi^2$=1.02).
The main spectral parameters are summarised in Table \ref{bb_best_fit}, while
Figure \ref{bb_eufspec} shows the fitted spectrum of IGR J14488--4008; note that 
in this case the cross calibration constants are closer to unity as listed in Table 3).

The source 20--100\,keV luminosity is $\sim$3$\times$10$^{44}$ erg/s.
Given the source black hole mass (M$_{\rm BH}$=3.8$\times$10$^{8}$M$_{\odot}$; \citealt{masetti:2013})
and employing the relation used by \citet{Mushotzky:2008} 
and scaled to the 20--100\,keV ISGRI luminosity
(L$_{\rm Bol}$=25$\times$L$_{\rm 20-100keV}$),
we estimate an average bolometric luminosity of $\sim$6$\times$10$^{45}$ erg/s and
consequently an average Eddington ratio of 0.16, making IGR J14488--4008 a quite efficient accretor; 
the uncertainty on the black hole mass gives a small range (0.1--0.3) 
for the Eddington ratio which therefore reinforces our conclusion that IGR J14488--4008 is
an efficiently accreting source.
Even using the 2--10\,keV luminosity and adopting the bolometric correction for radio loud
AGN discussed by \citet{runnoe:2012} the Eddington ratio remains high (0.1).
\begin{table}
\centering
\small
\caption{0.3--110\,keV Best fit spectral parameters. The model employed is \texttt{const*wa$_{\rm g}$*zwa*zedge*zedge*swind1*(pexrav+zga)})}
\label{bb_best_fit}
\begin{tabular}{ll}
\hline
Cold Neutral Absorption   & N$_{\rm H}$=(0.17$^{+0.04}_{-0.04})$$\times$10$^{22}$cm$^{-2}$\\
O$_{\rm VII}$ Abs. Edge          & E=0.73\,keV (fixed)\\
                          & $\tau_{\rm max}$= 2.52$^{+0.41}_{-0.41}$\\
Fe$_{\rm XX}$ Abs. Edge         & E=0.96\,keV (fixed)\\
                          & $\tau_{\rm max}$=0.71$^{+0.36}_{-0.31}$\\
\texttt{SWIND1} Param.    & N$_{\rm H}$=(7.85$^{+3.04}_{-2.37}$)$\times$10$^{22}$cm$^{-2}$\\
                          & Log($\xi$)=2.44$^{+0.09}_{-0.10}$\,erg\,cm$^{-1}$\,s$^{-1}$\\
                          & $\sigma$=0.22$^{+0.13}_{-0.08}$\\
Continuum slope           & $\Gamma$=1.71$^{+0.16}_{-0.17}$\\
Cut-off Energy            & E$_{\rm c}$=67$^{+227}_{-36}$\,keV\\
Reflection                & R= 1 (fixed) \\
Iron Line                 & E=6.45$^{+0.05}_{-0.04}$\,keV \\
                          & $\sigma$=0.03\,keV (fixed) \\
                          & EW=93$^{+35}_{-34}$\,eV\\
Cross. Calib. Const.      & C$_{\rm BAT}$=0.96$^{+0.45}_{-0.34}$\\
                          & C$_{\rm ISGRI}$=0.71$^{+0.42}_{-0.30}$\\   
%20--100\,keV Flux (BAT)   & 8.17$\times$10$^{-12}$erg\,cm$^{-2}$\,s$^{-1}$\\
%20--100\,keV Flux (IBIS)  & 6.01$\times$10$^{-12}$erg\,cm$^{-2}$\,s$^{-1}$\\                                                          
\hline
\end{tabular}
\end{table}

\section{Discussion and conclusions}

IGR J14488--4008 belongs to a sample of 
radio galaxies detected by INTEGRAL and currently under study with the Giant Metrewave Radio
Telescope (GMRT).
The high resolution provided by the GMRT has allowed
a proper determination of the nature of IGR J14488--4008,
confirming that it is a newly discovered giant radio galaxy
(GRG) with a linear size exceeding 1.5\,Mpc and a possible
X--shaped morphology. Within the sample of radio galaxies 
detected by INTEGRAL,
IGR J14488--4008 is the second source to be classified as a new GRG (the other being IGR J17488--2338
\citealt{molina:2014}); this confirms that the hard X-ray selection is indeed a 
valuable tool to identify this peculiar class of radio galaxies, albeit this
might be the result of observing in this energy band the most powerful objects.
Both IGR J14488--4008 and IGR J17488--2338 appear to be characterised by high
accretion rates, but when comparing their X-ray characteristics, these sources are quite
different: while IGR J17488--2338 does not have a very complex spectrum (a part from 
a cold absorber partially covering the nucleus), IGR J14488--4008 is characterised by
the presence of a strong warm absorber, which is indeed a very peculiar feature.

At soft energies, more than half of the radio quiet Seyfert 1 galaxies show complex intrinsic
absorption and/or emission lines associated with the presence of photoionised gas, 
the so-called warm absorber (e.g. \citealt{blustin:2005}; \citealt{mckernan:2007}).
Up to very recently, their radio loud counterparts, i.e. broad line radio galaxies (BLRG), were not
expected to show signs of ionised absorption. This was ascribed to the
presence of a jet seen at small angles with respect to the line of sight that would
mask the signature of ionised gas with its Doppler-booster, non-thermal radiation.
However, with the advent of more sensitive X-ray observatories, this assumption has been proven
to be incorrect and there is now a handful of sources for which the presence of
a warm absorber has indeed been confirmed (e.g. 3C 382, 3C 390.3, 3C 445, \citealt{torresi:2012};
4C 74.26, \citealt{Ballantyne:2005a}). Within this framework,
IGR J14488--4008 can be considered as a new addition to this small but growing sample
of BLRG with complex absorption signatures in their X-ray spectra.
Our analysis has also shown that the warm absorber observed in IGR J14488--4008 has
similar characteristics to the warm absorbers seen in other BLRG: same column densities
and ionisation parameter (see \citealt{torresi:2012}), although the gas outflow velocities 
are slightly higher, probably due to the different model adopted here. 

From the available information it is possible to evaluate the location
of the warm gas in IGR J14488--4008, warning the reader of the many uncertainties
which effect the performed calculation (imperfect modelling, large errors on some parameters,
various adopted assumptions).
Following the reasoning of \citet{torresi:2012}, we estimate the
minimum and maximum distance of the warm absorber to be 
R$_{\rm min}$=(2GM)/v$^2$ and R$_{\rm max}$=L$_{\rm ion}$/($\xi$N$_{\rm H}$).
Since the ionising luminosity (L$_{\rm ion}$) derived in the range 1--1000 Rydberg (13.6\,eV--13.6\,keV)
from our best fit is 2.9$\times$10$^{44}$erg/s, while the other parameters 
required in the above formula are those reported in Table 2,
we estimate the location of the warm gas to be between 0.01 and 4.4 parsecs.
In IGR J14488--4008 the location of the  Broad Line Region (BLR) and the torus 
can be obtained following \citet{ghisellini:2008} and \citet{torresi:2012}.
Assuming L$_{\rm ion}$=L$_{\rm disk}$, we find R$_{\rm BLR}$=10$^{17}$L$_{\rm disk}^{1/2}$=0.017\,pc
and R$_{\rm torus}$=2.5$\times$10$^{18}$L$_{\rm disk}^{1/2}$=0.4\,pc, where L$_{\rm disk}$ is
expressed in units of 10$^{45}$erg/s. These values suggest that the bulk of the highly ionised
absorbing gas detected in IGR J14488--4008 is located internally to the torus. We also note that
among BLRG with a warm absorber component, only for 3C 445 a similar location is reported
\citep{torresi:2012}, while in other objects the warm gas is found beyond the torus.
IGR J14488--4008 is also similar to 3C 445 for the high outflow mass. 

In summary, IGR J14488--4008 is a remarkable AGN; firstly discovered to be a high energy emitter by INTEGRAL,
it has been identified as a Seyfert 1.2 galaxy thanks to an optical follow-up study and
then further characterised as a Giant Radio Galaxy by means of a dedicated programme which is being 
carried out at the GMRT facility. IGR J14488--4008 appears to be a highly efficient accretor, 
and  is also peculiar in the X-ray domain,
being one of the few broad line radio galaxies known to have a warm and highly ionising gas
in its nuclear regions. Unlike other BLRG with warm absorbers, the warm gas 
in IGR J14488-4008 is located well within the
torus region, making it similar to only another BLRG, 3C 445.

\section*{Acknowledgements}
We acknowledge financial support from ASI under contract INTEGRAL
2013.23.RO. We thank the staff of the GMRT that made these observations
possible. GMRT is run by the National Centre for Radio Astrophysics
of the Tata Institute of Fundamental Research.

\bibliography{mol_biblio}

\end{document}